\documentclass[12pt]{amsart}
\usepackage{amsmath}
\usepackage{amssymb}
\usepackage[latin1]{inputenc}
\usepackage[T1]{fontenc}
\newcommand{\tr}{{\rm Tr }}

\newcommand{\bra}{\langle}
\newcommand{\ket}{\rangle}
\newcommand{\vp}{\varphi}

\newcommand{\N}{\mathbb{N}}

\newcommand{\R}{\mathbb{R}}
\newcommand{\be}{\begin{equation}}
\newcommand{\eeq}{\end{equation}}
\newcommand{\bet}{\begin{equation*}}
\newcommand{\eeqt}{\end{equation*}}
\newcommand{\bea}{\begin{eqnarray}}
\newcommand{\eeqa}{\end{eqnarray}}
\newcommand{\beat}{\begin{eqnarray*}}
\newcommand{\eeqat}{\end{eqnarray*}}

\newcommand{\hil}{\mathcal{H}}

\begin{document}
\title{Measuring position and momentum together}

\author{Paul Busch}
\address{Department of Mathematics, University of York,
York, UK}
\email{pb516@york.ac.uk}

\author{Jukka Kiukas}
\address{Department of Physics, University of Turku, Turku, Finland}
\email{jukka.kiukas@utu.fi}
\author{Pekka Lahti}
\address{Department of Physics, University of Turku, Turku, Finland}
\email{pekka.lahti@utu.fi}

\begin{abstract}
We describe an operational scheme for determining both the position and momentum
distributions in a large class of quantum states, together with an experimental implementation.\\
PACS numbers: 03.65.-w, 03.65.Ta, 03.65.Wj, 03.67.-a
\end{abstract}

\maketitle

The basic understanding of quantum mechanics asserts that due to their noncommutativity,
position and momentum of a quantum object cannot be measured together in any state. 
It is an equally well known fact that these observables can be {\em approximately} measured together  if the measurement accuracies fulfill an appropriate trade-off relation. Much effort has gone into obtaining rigorous formulations of these facts; for a recent overview see, for instance, \cite{BHL2007}.

In this paper we wish to point out that there are measurement schemes which allow one to determine both the position and the momentum distributions from the statistics of a \emph{single} measurement: in this sense the position and momentum of a quantum object can, after all, be measured together.

Technically, this observation builds on the well-known properties of some covariant phase space observables, such as the Husimi Q-function. The measurement in question can then be implemented via eight-port homodyne detection. Therefore, the idea is completely realistic.


The structure of covariant phase space observables is well-known.
If $S$ is a positive operator of trace one (acting on the Hilbert space $\hil$ of the quantum object), 
and  $W_{qp}$, $(q,p)\in\R^2$, are the Weyl operators  associated  with the position 
and momentum operators $Q$ and $P$, then the operators
$$
G^{S}(Z)=\frac 1{2\pi}\int_Z W_{qp}\,S\,W_{qp}^*dqdp,
$$
(with the integral being defined in the sense of the weak operator topology) constitute a covariant phase space observable $G^S$ (defined on the Borel subsets of the phase space  $\R^2$).
Conversely, any covariant phase space observable is of that form for some $S$ \cite{note1}.
If the generating operator $S$ is the vacuum state ${|0\ket}$, then the probability distribution  $Z\mapsto G_\rho^S(Z)\equiv\tr[\rho G^S(Z)]= \tr[\rho G^{|0\ket}(Z)]$ is 
 the Husimi distribution of the state (density operator) $\rho$.

We are going to show that for a certain family of operators $S$ the phase space observable $G^S$ entails complete information about the position and momentum observables: that is, the spectral measures of $Q$ and $P$ can be determined from $G^S$. This is done by using the moment operators $L(x^k,G^S)$ and $L(y^k,G^S)$ of $G^S$, which are defined as follows
(using the general theory of operator integrals):
\begin{eqnarray*}
L(x^k,G^S)&=&\frac 1{2\pi}\int_{\R^2}q^k\, W_{qp}\,S\,W_{qp}^*dqdp,\\
L(y^k,G^S)&=&\frac 1{2\pi}\int_{\R^2}p^k\, W_{qp}\,S\,W_{qp}^*dqdp.
\end{eqnarray*}
In general, it is very difficult to determine the domains of
these (unbounded) operators. However, the following is known \cite[Theorem 4]{KLY2006} (for a similar result, see \cite{Werner86}):
for any $k\in\N$, if (and only if) the operator $Q^k\sqrt{S}$ is a Hilbert-Schmidt operator, that is, 
$\tr[SQ^{2k}]<\infty$, then 
\begin{equation}\label{positionk}
L(x^k,G^S)=\sum_{l=0}^k s^Q_{kl}Q^l, \ {\rm with}\ s^Q_{kl}=\textstyle\binom kl (-1)^{k-l}\tr[Q^{k-l}S],
\end{equation}  
and, similarly, 
 if (and only if) the operator $P^k\sqrt{S}$ is a Hilbert-Schmidt operator, then 
\begin{equation}\label{momentumk}
L(y^k,G^S)=\sum_{l=0}^k s^P_{kl}P^l, \ {\rm with}\ s^P_{kl}=\textstyle\binom kl (-1)^{k-l}\tr[P^{k-l}S].
\end{equation}  

For any vector state $\vp$, the numbers
$\langle\vp|L(x^k,G^{S})\vp \rangle$ are the $k^{th}$ moments
of the marginal probability
measure $X\mapsto \langle \vp|G^{S}(X\times \R)\vp\rangle$, and
similarly for the numbers $\langle\vp|L(y^k,G^{S})\vp \rangle$.
From Eq~(\ref{positionk}) one can easily determine any $Q^k$ in terms of the (operationally well defined) moment operators $L(x^l,G^S)$, $l=0,\ldots k$,
and, similarly, each  $P^k$ in terms of the  (operationally well defined)
moment operators $L(y^l,G^S)$, $l=0,\ldots k$, from (\ref{momentumk}).
More explicitly, consider the simple recursive transformation
\begin{equation}\label{recursion}
\beta_k = \alpha_k -\sum_{l=0}^{k-1}s_{kl}\beta_l, \ \ k\in \N.
\end{equation}
Putting $\alpha_k = \langle\vp|L(x^k,G^{S})\vp \rangle$ and
$s_{kl}=s_{kl}^Q$ yields $\beta_k = \bra \vp|Q^k\vp\ket$. (Note that
$s^Q_{kk}=1$.)
Similarly, one gets the moments $\bra\vp |P^k\vp\ket$ from
the actual marginal moments $\langle\vp|L(y^k,G^{S})\vp \rangle$.
Therefore, the numbers $\bra \vp|Q^k\vp\ket$ and $\bra\vp |P^k\vp\ket$ are
just as operational as the actually measured moments
$\langle\vp|L(x^k,G^{S})\vp \rangle$ and $\langle\vp|L(y^k,G^{S})\vp \rangle$, from which they can be reconstructed via \eqref{recursion}. But $\bra \vp|Q^k\vp\ket$ and $\bra\vp |P^k\vp\ket$
are the moments of the probability distributions $p_\vp^Q$ and $p_\vp^P$
of the position and momentum observables $Q$ and $P$
in the initial state $\vp$.
If $\vp$ is chosen to be, for example, a linear combination of Hermite functions,
the measures $p_\vp^Q$ and $p_\vp^P$ are exponentially bounded, and, as such,
are uniquely determined by
their respective moment sequences $(\langle\vp|Q^k\vp\rangle)_k$ and $(\langle\vp|P^k\vp\rangle)_k$
\cite[p. 406, Theorem 30.1]{BillingsleyII}. \emph{In this sense
one is  able to measure the noncommuting observables $Q$ and $P$
simultaneously in such a vector state $\vp$.} Furthermore,
since the linear combinations of Hermite functions are dense in $L^2(\R)$, 
their associated distributions $p^Q_\vp$ and $p^P_\vp$
suffice to determine the whole position and momentum observables $Q$ and $P$ (as spectral measures). 

The phase space observables $G^S$, with the generating operator $S$ being such that $\tr[W_{qp}S]\ne 0$ for almost all $(q,p)\in\R^2$, are informationally complete: that is,
the statistics $G^S_\rho$ determine the state $\rho$ of the quantum object. Given  the state $\rho$, the distributions of all observables are then determined, including, in particular, $p^Q_\rho$ and $p^P_\rho$. 
In this sense one might argue that an informationally complete measurement serves as a measurement of any other observable of the system.
However, to our knowledge, there is no simple method  to determine from the statistics $G^S_\rho$ the statistics of another observable. An explicit reconstruction
of the state typically requires \emph{several different} measurements, as is the case with the usual tomographic methods.

In the present case we are able to construct directly and in simple operational  terms (without any use of the informational completeness)
the moments of the distributions $p^Q_\vp$ and $p^P_\vp$ from the statistics $G^{S}_\vp$ for any state $\vp$, and these moment sequences identify, in turn, the distributions themselves in the exponentially bounded case.

It is useful to note that  the moment sequence of, say, the momentum distribution does not, in general, identify that distribution, even in cases where all the moments are finite. For example, if $\vp_\delta=(\vp_1+e^{i\delta}\vp_2)/\sqrt 2$ is a {\em double-slit state}, that is, $\vp_1,\vp_2$ are  (smooth) wave functions supported on disjoint intervals, then the moments $\bra \vp_\delta |P^k \vp_\delta\ket$ of the momentum distribution $p^P_{\vp_\delta}$ are all finite and independent of $\delta$ although the distribution does depend on $\delta$. This highlights the fact that the assumption of exponential boundedness is crucial in the above consideration. 

We also want to point out that the exponential boundedness
condition itself is operational, since it can, in principle, be verified
by considering the very moment sequence that is obtained in the measurement.
In fact, a probability measure with the moment sequence $m_1,m_2,\ldots$ is
exponentially bounded if and only if
\begin{equation}\label{expbound}
|m_k|\leq C R^k k!,  \ \ k=1,2,\ldots,
\end{equation}
for some $C,R>0$ (see e.g. the proof of Proposition 2 of \cite{Simon}).

Experimental implementations of observables $G^S$ can be obtained in quantum optics. For example,
the Husimi Q-function of a single-mode signal field can actually be measured by the eight-port homodyne detection technique, provided that the reference beam (used for the homodyne detection) is a very strong coherent field so that it can be treated classically; see e.g. \cite{Leonhardt_Paul}.
In a recent paper it has been shown, {\em without any classicality assumptions},  that {\em any} covariant phase space observable $G^S$ can be realized as a high-amplitude
limit of an eight-port homodyne detection observable choosing the auxiliary parameter field to be in the conjugated state $C^{-1}SC$ \cite{KL2007b}.

In this measurement, the input signal field, in state $\rho$, is mixed by a beam
splitter with another single-mode field, the latter being in the reference state $C^{-1}SC$.
Each of the two output beams of the beam splitter are led to a
homodyne detector which measures a field quadrature. One measures
$\frac{1}{\sqrt{2}}(a^*+a)$ and the other one
$\frac{1}{\sqrt{2}}i(b^*-b)$, where $a$ and $b$ are
the annihilation operators for the signal and the reference field
(in the Schr\"odinger picture).
When this measurement is interpreted with respect to the signal field in the
initial state
$\rho$, the combined statistics of the homodyne detectors are given by the
distribution
$(X,Y)\mapsto \tr[\rho G^{S}(X\times Y)]$, where the Weyl operators $W_{qp}$
are now defined in terms of the signal field quadratures
$Q:=\frac{1}{\sqrt 2}(a^*+a)$
and $P:=\frac {1}{\sqrt 2}i(a^*-a)$.

By performing the measurement, one thus obtains two outcomes, which then are to be interpreted with respect to the signal beam.
Notice that the outcomes for the two homodyne detectors are obtained at the same time. Consequently, upon repeating the preparation and the measurement,
the statistics for the detectors are likewise created simultaneously.
After a suitable statistical accuracy has been obtained, and the
measurement is thus complete, one takes the statistics given by,
say, detector 1, computes its moments $\alpha_k$,
and calculates the numbers $\beta_k$ using \eqref{recursion}. The $\beta_k$ are
now the moments of the distribution of $Q$, in the input state of the signal
field. A similar procedure applied to detector 2 yields the moments $\beta'_k$
of the distribution of $P$. One can then check whether \eqref{expbound}
holds for both the sequences $\beta_k$ and $\beta_k'$. If that is the
case, then it is justified to say that (the statistics of) the position- and momentum-like
observables $Q$ and $P$ have been measured together. 

\

{\bf Acknowledgment.} One of us (J. K.) was supported by Emil Aaltonen
Foundation during the preparation of the manuscript.


\begin{thebibliography}{99}
\bibitem{note1} This result goes back to Holevo \cite{Holevo} and
Werner \cite{WernerII}, detailed proofs with different techniques are given in \cite{Cassinelli} and \cite{wernerarticle}.
\end{thebibliography}
\end{document}